\documentclass[aps,preprint,amsmath,amssymb,epsfig]{revtex4}
\usepackage{graphicx}% Include figure files
\usepackage{bm}% bold math
\usepackage{dcolumn}% Align table columns on decimal point
\usepackage{ulem}
\usepackage{xfrac}
\usepackage[printwatermark]{xwatermark}
\usepackage{xcolor}
\usepackage{graphicx,amsmath,amssymb}
\usepackage{epstopdf}
\usepackage{placeins}
\usepackage[colorlinks=true, urlcolor=blue, linkcolor=blue, citecolor=blue]{hyperref}

\begin{document}

\title{Experimental determination of the QCD effective charge $\alpha_{g_1}(Q)$}

\author{
A. Deur,$^{\njlab, \nuva}$
V. Burkert,$^{\njlab}$
J.P. Chen,$^{\njlab}$
W. Korsch$^{\nuk}$ 
}

\affiliation{
\centerline{{$^{\njlab}$Thomas Jefferson National Accelerator Facility, Newport
News, VA 23606. USA}}
\centerline{{$^{\nuva}$ University of Virginia, Charlottesville, VA 22904. USA}}
\centerline{{$^{\nuk}$University of Kentucky, Lexington, KY 40506. USA}}
}

\newcommand{\njlab}{1}
\newcommand{\nuva}{2}
\newcommand{\nuk}{3}

\begin{abstract}
The QCD effective charge  $\alpha_{g_1}(Q)$ is an observable that characterizes the magnitude of the strong interaction.
At high momentum $Q$, it coincides with the QCD running coupling $\alpha_{\rm s}(Q)$.
At low $Q$, it offers a nonperturbative definition of the running coupling.
We have extracted $\alpha_{g_1}(Q)$ from measurements
carried out at Jefferson Lab that span the very low to moderately high $Q$ domain, 
$0.14 \leq Q \leq 2.18$~GeV. 
The precision of the new results is much improved over the previous extractions and the reach in $Q$ at the lower end is significantly expanded.  
The data show that $\alpha_{g_1}(Q)$ becomes $Q$-independent  at very low $Q$. 
They compare well with two recent predictions of the QCD effective charge based on Dyson-Schwinger equations
and on the AdS/CFT duality.
\end{abstract}

\maketitle

\section{Introduction}
The behavior of Quantum Chromodynamics (QCD), the gauge theory of the strong interaction, 
is determined by the magnitude of its coupling $\alpha_{\rm s}$. It is large at low momentum, characterized here by  
$Q \equiv \sqrt{-q^2}$ with $q^2$ the square of momentum transferred in the process of electromagnetically probing a hadron. 
For $Q \ll 1$~GeV, $\alpha_{\rm s} (Q)  \gtrsim 1$ which is one of the crucial pieces leading to quark confinement. For $Q \gg1$~GeV, 
$\alpha_{\rm s} (Q) \lesssim 0.2$ which enables the use of perturbative computational techniques  (perturbative QCD, pQCD) constituting an accurate analytical approximation of QCD. In this domain, $\alpha^{\rm pQCD}_{\rm s}$ is well defined and known within an accuracy of 1\%  at $Q=M_{Z^0}=91$~GeV, the $Z^0$ mass, and within a few percents at $Q$ values of a few GeV~\cite{dEnterria:2022hzv}.
However, using  pQCD at $Q \lesssim 1$ GeV produces a diverging $\alpha^{\rm pQCD}_{\rm s}$ (Landau pole) that prohibits  
any perturbative expansion in $\alpha^{\rm pQCD}_{\rm s}$ and signals the breakdown of pQCD. 
In contrast, most nonperturbative methods, including lattice QCD~\cite{Zyla:2020zbs}, the 
AdS/CFT (Anti-de-Sitter/Conformal Field Theory) duality~\cite{Brodsky:2014yha, Dobado:2019fxe} 
implemented  using QCD's light-front (LF) quantization~\cite{Dirac:1949cp} and a soft-wall AdS potential (Holographic LF QCD, HLFQCD~\cite{Brodsky:2003px})
or solving the Dyson-Schwinger equations (DSEs)~\cite{Maris:2003vk} yield a finite
$\alpha_{\rm s}$. In fact, many theoretical approaches predict that $\alpha_{\rm s}$ 
``freezes'' as $Q \to 0$, {\it viz}, it loses its $Q$-dependence~\cite{Deur:2016tte}. 

There are several possible definitions of $\alpha_{\rm s}$ in the nonperturbative domain  ($Q \lesssim 1$~GeV)~\cite{Deur:2016tte}. 
We use here the {\it effective charge} approach that defines 
$\alpha_{\rm s}$  from the perturbative series of an observable truncated to its first order in $\alpha_{\rm s}$~\cite{Grunberg:1980ja}. 
Although this definition can be applied for any $Q$ value, it was initially proposed for the pQCD domain where it 
makes $\alpha_{\rm s}$ the equivalent of the Gell-Mann Low coupling of Quantum Electrodynamics (QED), $\alpha$~\cite{GellMann:1954fq}. 
With this definition, $\alpha_{\rm s}$ can be evaluated at any $Q$ value, has no low $Q$ divergence
and is analytic around quark mass thresholds. Furthermore, since
the first order in $\alpha^{\rm pQCD}_{\rm s}$ of a pQCD approximant is independent of the choice of renormalization 
scheme (RS), effective charges are independent of RS and gauge choices. 
This promotes $\alpha_{\rm s}$ from a parameter depending on chosen 
conventions to an observable, albeit with the caveat that it becomes process-dependent since 
two observables produce generally different 
effective charges. Yet, pQCD predictability is maintained because effective charges are related without renormalization scale ambiguities by
Commensurate Scale Relations (CSR)~\cite{Brodsky:1994eh}.
CSR are known to hold for pQCD and QED since the latter corresponds to the 
$N_C\to 0$ limit of QCD, with $N_C$ the number of colors.
For example, CSR explicitly relate $\alpha_{g_1}$, $\alpha_{F_3}$, $\alpha_{\tau}$ and $\alpha_R$ defined using the generalized Bjorken sum rule~\cite{Bjorken:1966jh}, the Gross-Llewellyn Smith sum rule~\cite{Gross:1969jf}, and the perturbative approximant for the $\tau$-decay rate~\cite{Brodsky:2002nb} and $R_{e^+e-}$~\cite{Gorishnii:1990vf}, respectively.
In fact, the choice of process to define an effective charge is analogous 
to a RS choice for $\alpha^{\rm pQCD}_{\rm s}$~\cite{Deur:2014qfa} and the procedure of
extracting an effective charge, e.g., from $\tau$-decay is denoted the $\tau$-scheme. 
Here, we discuss the effective charge $\alpha_{g_1}(Q)$ ($g_1$-scheme) extracted using the generalized Bjorken sum rule: 
\begin{eqnarray}
\Gamma_1^{\rm p-n}(Q^2 )  \equiv  \int_0^{1^-} g_1^{\rm p}(x,Q^2)- g_1^{\rm n}(x,Q^2) dx = \frac{g_{\rm A}}{6}\bigg[1-\frac{\alpha^{\rm pQCD}_{{\rm {s}}}(Q)} 
{\pi}-3.58\left(\frac{\alpha^{\rm pQCD}_{{\rm {s}}}(Q)}{\pi}\right)^2  \nonumber \\
-20.21\left(\frac{\alpha^{\rm pQCD}_{{\rm {s}}}(Q)}{\pi}\right)^{3} -
175.7\left(\frac{\alpha^{\rm pQCD}_{{\rm {s}}}(Q)}{\pi}\right)^{4}+\mathcal O\left(\big(\alpha^{\rm pQCD}_{\rm {s}} \big)^5\right)... \bigg]+\sum_{n > 1} \frac{\mu_{2n}}{Q^{2n-2}},
\label{eq:genBj}
\end{eqnarray}
where $x$ is the Bjorken scaling variable~\cite{Bjorken:1968dy}, 
$g_{\rm A}=1.2762(5)$~\cite{Zyla:2020zbs} is the nucleon axial charge,
$ g_1^{\rm p(n)}$ is the longitudinal spin structure function of the proton(neutron) obtained in polarized 
lepton-nucleon scattering~\cite{Deur:2018roz}
and $\mu_{2n}$ are the Operator Product Expansion's (OPE) nonperturbative higher twist (HT) terms. 
The integral excludes the elastic contribution at $x=1$. The series 
coefficients are computed for $n_f=3$ and in the $\overline{\rm MS}$ RS for 
the $n>1$ $\alpha_{\rm {s}}^n$ terms~\cite{Kataev:1994gd}.
They originate from the pQCD radiative corrections. Although the expansion~(\ref{eq:genBj}) is only applicable in the perturbative domain, i.e., at distance scales where confinement effects are weak, the HT terms can be related to the latter~\cite{Burkardt:2008ps} and one may picture the terms of Eq.~(\ref{eq:genBj}) as coherently merging together at low $Q$ to produce confinement.  

 The effective charge $\alpha_{g_1}$ is defined from 
 Eq.~(\ref{eq:genBj}) expressed at first order in coupling and twist: 
\begin{eqnarray}
\Gamma_1^{\rm p-n}(Q^2 ) \equiv \frac{g_{\rm A}}{6}\left(1-\frac{\alpha_{g_1}(Q)} 
{\pi} \right)
~\longrightarrow~
\alpha_{g_1}(Q) \equiv \pi \left(1-\frac{6}{g_{\rm A}}\Gamma_{1}^{\rm p-n}(Q) \right).
\label{eqn:alphadef}
\end{eqnarray}
Thus, in the domain where Eq.~(\ref{eqn:alphadef}) applies, $\alpha_{g_1}$ can be interpreted as a running coupling that not only includes short-distance effects such as vertex correction and vacuum polarization, but all other effects, e.g., pQCD radiative corrections and, in the lower-$Q$ domain of pQCD, HT terms
and other nonperturbative effects not formalized by OPE and therefore not included in Eq.~(\ref{eqn:alphadef}). The latter comes from coherent reactions of a
hadron (resonances). In the nonperturbative domain where pQCD radiative corrections and HT effects have merged into global confinement effects, $\alpha_{g_1}$ may approximately retain its
interpretation as a coupling if the contribution to $\Gamma_{1}^{\rm p-n}$ of nonresonant reactions continues to
dominate, as they do at large $Q$~\cite{Deur:2009zy}.

There are several advantages to $\alpha_{g_1}$~\cite{Deur:2016tte}. 
First, rigorous sum rules constrain  $\alpha_{g_{1}}(Q)$ for $Q \to 0$ (the Gerasimov--Drell--Hearn 
(GDH) sum rule~\cite{Gerasimov:1965et}) 
and $Q \to \infty$ (the Bjorken sum rule). They provide analytical expressions of $\alpha_{g_{1}}(Q)$ 
in these limits (blue dashed line and cyan hatched band in Fig.~\ref{fig:alpha}). 
Furthermore, contributions from $\Delta$ baryons are quenched in $\Gamma_{1}^{\rm p-n}$~\cite{Burkert:2000qm},
enhancing the nonresonant reactions contribution to $\Gamma_{1}^{\rm p-n}$ relatively to
the resonance contribution, which helps toward interpreting $\alpha_{g_1}$ as a coupling.
If so, $\alpha_{g_1}$ would remain approximately equivalent to the Gell-Mann Low coupling 
in the nonperturbative domain, a crucial property that it is not obvious and may be
specific to $\alpha_{g_1}$. Such property is supported by the agreement between $\alpha_{g_1}$
and calculations of couplings~\cite{Brodsky:2010ur, Binosi:2016nme} using a definition consistent with $\alpha_{g_1}$.

Former extractions of $\alpha_{g_1}$~\cite{Deur:2005cf}  were obtained from experimental data on $\Gamma_{1}^{\rm p-n}$ 
from CERN~\cite{Adeva:1993km}, DESY~\cite{Airapetian:2000yk}, Jefferson Lab (JLab)~\cite{Deur:2004ti} 
and SLAC~\cite{Anthony:1996mw}, see Fig.~\ref{fig:alpha}.
Since the results reported in Ref.~\cite{Deur:2005cf}, progress occurred  on both the experimental and 
theoretical fronts.
Firstly, when Ref.~\cite{Deur:2005cf} was published, the meaning of $\alpha_{g_1}$ in the nonperturbative 
region was unclear. Thus, the comparison in~\cite{Deur:2005cf} of $\alpha_{g_1}$ to theoretical predictions of the
nonperturbative coupling was tentative. 
This is now better understood: as just discussed, $\alpha_{g_1}$ essentially  retains its meaning of effective charge at low 
$Q$~\cite{Deur:2009zy, Deur:2016tte}. 
Secondly, new data on $\Gamma_{1}^{\rm p-n}$ have become available from 
CERN (COMPASS experiment)~\cite{Alekseev:2010hc} and
JLab (EG1dvcs experiment)~\cite{Deur:2014vea}  at high $Q$, and from JLab (E97110, E03006 and E05111 
experiments)~\cite{Deur:2021klh} at very low $Q$. 
Finally, new theoretical studies of the nonperturbative behavior of $\alpha_{\rm s}$ were conducted, 
including the first use of the AdS/CFT duality to describe the strong coupling in its nonperturbative domain~\cite{Brodsky:2010ur} and
the identification of a process-independent (PI) effective charge $\hat \alpha_{\rm PI}(Q)$ 
that unifies a large body of research 
from DSE and lattice QCD to $\alpha_s$~\cite{Binosi:2016nme, Rodriguez-Quintero:2018wma}. 
Connections between the nonperturbative and 
perturbative effective charges were made~\cite{Deur:2014qfa, Deur:2017cvd, Deur:2016tte}, 
which permitted a prediction of $\alpha_{\rm s}$ at the $Z_0$ pole,
$\alpha_{\rm s}^{\overline{\rm MS}}(M_z^2)=0.1190\pm0.0006$ at N$^3$LO~\cite{Deur:2016opc} 
that agrees well with the 2021 Particle Data Group compilation, $\alpha_{\rm s} (M_{\rm Z}) = 0.1179\pm0.0009$~\cite{Zyla:2020zbs}.
In addition to predicting quantities characterizing hadronic structures~\cite{Brodsky:2014yha, Binosi:2016nme, Cui:2020tdf}, 
the effective charge helps establish conformal behavior at
low $Q$. Through AdS/CFT, this helps the investigation of
the physics beyond the standard model~\cite{Dobado:2019fxe} or 
of the quark-gluon plasma~\cite{Janik:2010we} in heavy ion collisions~\cite{Busza:2018rrf} and nuclear hydrodynamics~\cite{Florkowski:2017olj} for the latter and neutron stars~\cite{Jokela:2018ers}.
\\
Here, we report on new experimental data on $\alpha_{g_1}$ extracted from~\cite{Alekseev:2010hc, Deur:2014vea, Deur:2021klh}  and how they compare with the latest theory predictions.

\section{Experimental extraction of $\alpha_{g_1}$}
The new JLab data on $\Gamma_{1}^{\rm p-n}(Q)$ were taken by four experiments. 
The first experiment, E97110~\cite{Sulkosky:2019zmn}, occurred in the Hall A~\cite{Alcorn:2004sb}  of JLab.
The three others used the CLAS spectrometer~\cite{CLAS:2003umf} in JLab's Hall B and were experiments EG1dvcs~\cite{Prok:2014ltt}, 
E03006~\cite{Zheng:2021yrn} and E05111~\cite{Adhikari:2017wox} (the two latter being referred to as Experimental Group EG4).
The four experiments occurred  during the 6 GeV era of JLab, before its 12 GeV upgrade.  
The experiments used a polarized electron beam with energies ranging from 0.8 to 6 GeV.
E97110 studied the spin structures of the neutron and $^3$He using the 
Hall A  polarized $^3$He target with longitudinal and transverse polarization directions~\cite{VINCETHESIS}.
EG1dvcs, E03006, E05111 studied the proton, neutron  and deuteron spin structures using the Hall B longitudinally 
polarized ammonia (NH$_3$ or ND$_3$) target~\cite{Keith:2003ca}. 
The main purpose of EG1dvcs was high $Q$, up to 2.65 GeV ($Q^2=7$~GeV$^2$), exclusive measurements of 
Deep Virtual Compton Scattering. 
Therefore, it provided highly precise inclusive $\Gamma_{1}^{\rm p-n}$ data compared to the older data
in the same domain~\cite{Adeva:1993km, Anthony:1996mw, Airapetian:2000yk, Deur:2004ti}. 
E97110, E03006 and E05111 were dedicated to test Chiral Effective Field Theory %($\chi$EFT) 
predictions by covering very low $Q$ domains: $0.19 \leq Q \leq 0.49 $, $0.11 \leq Q \leq 0.92$  and 
$0.14 \leq Q \leq 0.70 $~GeV, respectively. 
To reach low $Q$ while covering the large $x$ range necessary for the $\Gamma_{1}$ integral, 
high beam energy (up to 4.4 GeV) was needed and the scattered electrons 
had to be detected at small angles
(down to about $5^\circ$). In Hall A, the low angles were reached {\it via} a supplementary dipole magnet 
installed in front of the spectrometer~\cite{septum}. In Hall B, a Cherenkov Counter designed for high efficiency
at small angle was installed in one of the six sectors of CLAS~\cite{Adhikari:2017wox} which magnetic field was set 
to bent outward the scattered electrons. 
In addition, both the Hall A and B targets were placed about 1~m upstream of their usual positions. 

\begin{figure}[ht!]
\begin{center}
\centerline{\includegraphics[width=0.5\textwidth, angle=0]{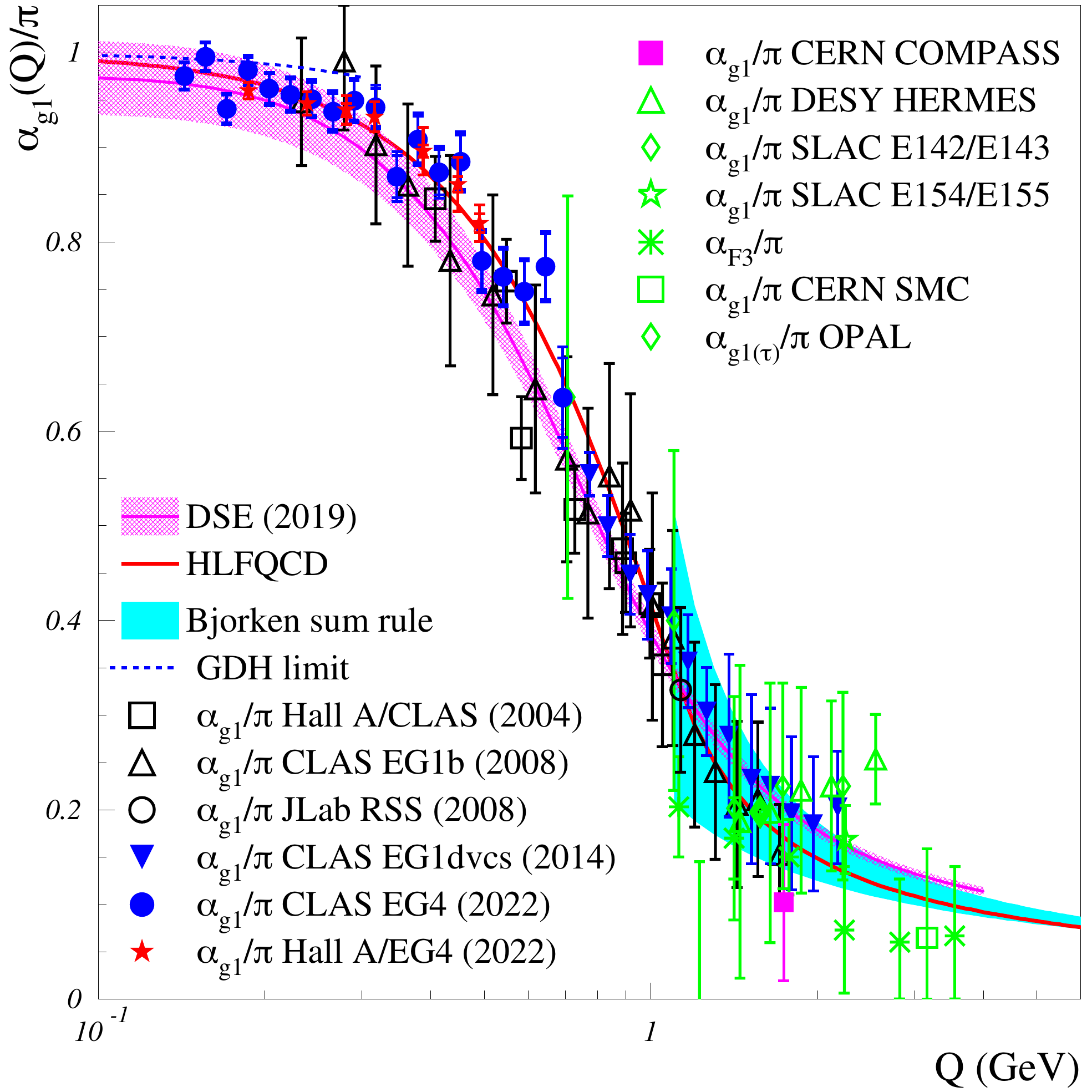}}
\end{center}
\caption{\footnotesize Effective charge $\alpha_{g_1}(Q)/\pi$ obtained from JLab experiments E03006/E97110~\cite{Deur:2021klh} (solid stars), E03006/E05111~\cite{Deur:2021klh} 
(solid circles) and EG1dvcs~\cite{Deur:2014vea}   (solid triangles) and from COMPASS~\cite{Alekseev:2010hc} (solid square). 
Inner error bars represent the statistical uncertainties and outer ones the systematic and statistical 
uncertainties added quadratically. 
The open symbols show the older world data~\cite{Adeva:1993km, Anthony:1996mw, Airapetian:2000yk, Deur:2004ti}
with the error bars the quadratic sum of the systematic and statistical uncertainties. 
Also shown are the HLFQCD~\cite{Brodsky:2010ur} (red line, using the HLFQCD scale $\kappa=0.534$~GeV~\cite{Sufian:2018cpj}) and DSE~\cite{Binosi:2016nme} 
(magenta line and hatched band) parameter-free predictions of effective charges. 
The dashed line and hatched cyan band are  $\alpha_{g_1}(Q)/\pi$ obtained from 
the GDH and Bjorken sum rules, respectively.}
\label{fig:alpha}
\end{figure}

The EG1dvcs data on proton and deuteron were combined to form $\Gamma_{1}^{\rm p-n}$ over the range 
$0.78 \leq Q \leq 2.18 $~GeV~\cite{Deur:2014vea}. The $\Gamma_{1}^{\rm p-n}$ formed with the E97110 and EG4, 
data  covers the $0.14 \leq Q \leq 0.70$~GeV range~\cite{Deur:2021klh}.
The $\alpha_{g_1}$ data, obtained 
following Eq.~(\ref{eqn:alphadef}), are shown in Fig.~\ref{fig:alpha} and given in Table~\ref{tab:alpha}. 
Also shown in the figure are the older data presented in Ref.~\cite{Deur:2004ti} 
including $\alpha_{F_3}$ extracted from the data~\cite{Kim:1998kia}  
and $\alpha_{g_1(\tau)}$ from the OPAL data on $\tau$-decay~\cite{Brodsky:2002nb}. 
The effective charge $\alpha_{F_3}$ is nearly identical
to $\alpha_{g_1}$~\cite{Deur:2005cf}, and $\alpha_{g_1(\tau)}$ was transformed from the $\tau$-scheme to the $g_1$-scheme using 
the CSR~\cite{Brodsky:1994eh}. Consequently, $\alpha_{F_3}$ and $\alpha_{g_1(\tau)}$ are directly comparable to $\alpha_{g_1}$.
We also show in Fig.~\ref{fig:alpha}  the theory predictions from AdS/CFT~\cite{Brodsky:2010ur}
and DSE~\cite{Binosi:2016nme}. Remarkably, both predictions are parameter-free and gauge-invariant.

The AdS/CFT coupling $\alpha^{\rm HLF}_{g_1}$ is obtained in the HLFQCD 
approach where QCD is quantized using LF coordinates~\cite{Dirac:1949cp}.  
The use of the HLFQCD approach incorporates the underlying conformal (i.e., scale-invariant) character of QCD at low and large $Q$. 
The deformation of the AdS$_5$ space is dual to a semiclassical potential that models quark confinement. 
This potential can be determined with various methods that all lead
to the same harmonic oscillator form~\cite{Brodsky:2014yha, deAlfaro:1976vlx, Trawinski:2014msa}. 
The effective charge $\alpha^{\rm HLF}_{g_1}$ is dual 
to the product of the AdS$_5$ coupling {\it constant} by the AdS$_5$ space deformation term. Since the 
latter is dual to the CFT confinement force, the meaning of 
$\alpha^{\rm HLF}_{g_1}$ is analogous to that of $\alpha_{g_1}$ which at low $Q$ incorporates in $\alpha_{\rm s}$ 
confinement effects. 
The $Q$-dependence of $\alpha^{\rm HLF}_{g_1}$ is controlled by a single scale, 
e.g., the proton mass. 
The coupling is normalized to 
$\alpha^{\rm HLF}_{g_1}(0)=\pi$ to obey the kinematic constraint that  $\Gamma_{1}^{\rm p-n}(0)=0$, i.e.,
$\alpha_{g_1}(0)=\pi$, see Eq.~(\ref{eqn:alphadef}). This normalization amounts to the RS choice of pQCD~\cite{Deur:2014qfa}. 
Thus, the $\alpha^{\rm HLF}_{g_1}(Q)$ prediction is parameter-free.
Above $Q \simeq 1$~GeV HLFQCD stops to be valid because its {\it semiclassical} potential 
does not include, by definition, the short distance quantum effects responsible for the running of a coupling. 
This is palliated by matching HLFQCD and pQCD near $Q\simeq 1$~GeV where both formalisms apply, thereby providing $\alpha^{\rm HLF}_{g_1}(Q)$ at all $Q$~\cite{Deur:2014qfa}. 

The DSE effective charge $\hat \alpha_{\rm PI}$~\cite{Binosi:2016nme} is obtained starting with the 
Pinch Technique~\cite{Cornwall:1981zr}  and Background Field Method~\cite{Abbott:1980hw}. 
They allow us to define a process-independent QCD coupling in terms of a mathematically reconstructed gluon two-point function analogous to the Gell-Mann Low effective charge of QED. 
The $\hat \alpha_{\rm PI}$ is then computed by combining the solution of DSE compatible with lattice QCD results.
The definition of $\hat \alpha_{\rm PI}$ explicitly factors in a renormalization group invariant interaction, thus making it, like
$\alpha_{g_1}(Q)$ and $\alpha^{\rm HLF}_{g_1}(Q)$, to incorporate confinement~\cite{Binosi:2014aea}.
Like them, $\hat \alpha_{\rm PI}(Q)$ freezes at low $Q$ with a predicted
infrared fixed-point of $\hat \alpha_{\rm PI}(0)=(0.97\pm 0.04)\pi$. 
The mechanism at the origin of the freezing in the DSE 
framework is the emergence of a dynamical gluon mass $m_g(Q)$~\cite{Cornwall:1981zr, Aguilar:2008xm} 
that (A) regulates the Landau pole and (B) decouples 
the dynamics at scales $Q 	\lesssim m_g(0)$, thereby causing the coupling to lose its 
$Q$-dependence~\cite{Brodsky:2008be}.  
Like $\alpha^{\rm HLF}_{g_1}$, $\hat \alpha_{\rm PI}$ is parameter-free and gauge-invariant but, in contrast to the former and $\alpha_{g_1}$, $\hat \alpha_{\rm PI}$ is also process-independent.
No parameter is varied to predict the infrared fixed-point $\hat \alpha_{\rm PI}(0)$ since it is largely fixed by the value of $m_g(0)$, nor a matching is necessary to ensure agreement with the perturbative determination of $\alpha^{\rm pQCD}_{\rm {g_1}}$ from the renormalization group equations
and the Bjorken sum rule. 
Crucially, the practical determination of $\hat \alpha_{\rm PI}(Q)$ 
consistently incorporates the extensive information from Lattice QCD on the gluon and ghost propagators, thereby 
connecting this technique to $\alpha_{g_1}$.

The new data on $\alpha_{g_1}$ agree well with the older data and display a much improved precision
over the whole $Q$ range covered. In addition, the data now reach clearly the freezing domain of QCD 
at very low $Q$. That $\alpha_{g_1}$ freezes could be already inferred with the old data but only by complementing them
with the GDH sum rule or/and the $\alpha_{g_1}(0)=\pi$ constraint. For the first time, the onset of freezing is now visible with data only.
One notes that only three of the lowest $Q$ points agree with the GDH expectation. This may signal  a fast arising
$Q$-dependence beyond the leading behavior given by GDH. The data agree well with the $\alpha^{\rm HLF}_{g_1}$ and
$\hat \alpha_{\rm PI}$ predictions. That such agreements would occur was not obvious and is a significant finding. 
The possible tension between the data and $\hat \alpha_{\rm PI}$ for the range $0.3 \lesssim Q \lesssim 0.5$~GeV may be because $\alpha_{g_1}$ and $\hat \alpha_{\rm PI}$ are not exactly the same 
effective charges (e.g., at high $Q$, 
$\sfrac{\alpha_{g_1}}{\hat \alpha_{\rm PI}}\simeq1+0.05\alpha^{\rm pQCD}_{\rm {s}} \neq 1$), but
it is noteworthy that it occurs only in the moderately low $Q$ domain where the ghost-gluon vacuum 
effect as computed in the Landau gauge contributes the most to $\hat \alpha_{\rm PI}$.

\section{Summary and conclusion}
We used the new JLab data and COMPASS datum on the Bjorken sum to extract the QCD effective 
charge $\alpha_{g_1}(Q)$ in the $Q$-range $0.14 \leq Q \leq 2.18 $~GeV. The new result displays a significantly higher precision compared to the older extractions of $\alpha_{g_1}(Q)$, and improve the low $Q$ reach by about a factor of 2. 

The new data show that $\alpha_{g_1}(Q)$ ``freezes'', {\it viz}, loses its $Q$-dependence, at small $Q$, saturating at 
an infrared fixed-point $\alpha_{g_1}(Q\simeq0) \simeq \pi$.
This was already apparent with the older data when combined with the GDH sum rule expectation, but the new data   
explicitly display the behavior without needing the sum rule and with significantly higher precision. 
The freezing of $\alpha_{g_1}(Q)$ together with the smallness of the light quark masses, makes
QCD approximately conformal at low $Q$. 
The conformal behavior vanishes when transiting from the low-$Q$ effective degrees of freedom of
QCD (hadrons) to the large-$Q$  fundamental ones (partons) where conformality is then restored (the long-known Bjorken scaling~\cite{Bjorken:1968dy}).  
This transition is revealed by the drastic change of value of the effective charge. It occurs at a $Q$ value indicative of the chiral symmetry breaking parameter, $\Lambda_B\simeq 1$~GeV. The breaking at low $Q$ of chiral symmetry, one of the crucial properties of QCD, is believed to cause the emergence of the global properties of hadrons.

The new data  agree well with sum rule predictions and with the latest predictions from DSE and from a AdS/CFT-based approach. 
It shows that a strong coupling can be 
consistently defined in the nonperturbative domain of QCD, namely as an effective charge analogous to the definition used in QED, and that it can then be used to compute a large variety of hadronic quantities and other phenomena in which the strong interaction plays a role.

~

\noindent {\bf Acknowledgments} 
The authors thank 
D. Binosi,  S. J. Brodsky, Z.-F. Cui, G. F. de T\'eramond, J. Papavassiliou, C. D. Roberts and J. Rodríguez-Quintero 
for their valuable comments on the manuscript. 
This work is supported by the U.S.\ Department of Energy, Office of Science, Office of Nuclear Physics, contracts DE-AC05-06OR23177 and DE-FG02-99ER41101.

%%%%%%%%%%%%%%%%%%%%%%%
\begin{table}[ht]
\centering
%EG4 
{\footnotesize
\begin{tabular}[t]{|c|c|}
\hline 
$Q$ (GeV) & $\alpha_{g_1} \pm \rm stat. \pm syst.$  \\
\hline 
\hline 

 0.143  & 3.064 $\pm 0.043 \pm 0.018 $ \\
\hline 
 0.156  & 3.129 $\pm 0.046 \pm  0.019 $ \\
\hline 
0.171  & 2.955 $\pm 0.046 \pm 0.023 $ \\
\hline 
0.187  & 3.083 $\pm 0.044 \pm 0.024 $ \\
\hline 
0.204  & 3.022 $\pm 0.049 \pm 0.024 $ \\
\hline 
0.223  & 3.002 $\pm 0.052 \pm 0.027 $ \\
\hline 
0.243  & 2.988 $\pm 0.055 \pm 0.031 $ \\
\hline 
0.266  & 2.947 $\pm 0.060 \pm 0.035 $ \\
\hline 
0.291  & 2.983 $\pm  0.065 \pm 0.035 $  \\
\hline 
0.317  & 2.961 $\pm 0.062 \pm 0.038 $ \\
\hline 
0.347  & 2.730 $\pm 0.070 \pm 0.044 $ \\
\hline 
0.379  & 2.853 $\pm 0.077 \pm 0.040 $ \\
\hline 
0.414  & 2.745 $\pm 0.076 \pm 0.041 $ \\
\hline 
0.452  & 2.779 $\pm 0.090 \pm 0.043 $ \\
\hline 
0.494  & 2.451 $\pm 0.094 \pm 0.044 $ \\
\hline 
0.540  & 2.397 $\pm 0.092 \pm 0.039 $ \\
\hline 
0.590  & 2.349 $\pm 0.101 \pm 0.040 $ \\
\hline 
 0.645  &  2.431 $\pm 0.109 \pm 0.043 $ \\
\hline 
0.704  & 1.996 $\pm 0.131 \pm 0.104 $ \\
\hline 
\end{tabular}
% Now E97110/EG4
\begin{tabular}[t]{|c|c|}
\hline 
$Q$ (GeV) & $\alpha_{g_1} \pm \rm stat. \pm syst.$  \\
\hline 
\hline 
 0.187 &  3.016 $\pm 0.009 \pm 0.027 $ \\
\hline
0.239 &  2.973 $\pm 0.015 \pm 0.035 $ \\
\hline
0.281 &  2.952 $\pm 0.021 \pm 0.041 $ \\
\hline
 0.316 &  2.929 $\pm 0.017 \pm 0.048 $ \\
\hline
0.387 & 2.815 $\pm 0.021 \pm 0.076 $  \\
\hline
0.447 &  2.704 $\pm 0.025 \pm 0.086 $ \\
\hline
0.490 &  2.575 $\pm 0.031 \pm 0.053 $ \\
\hline 
% now EG1dvcs
\hline 
0.775  & 1.743 $\pm 0.007 \pm 0.071 $ \\
\hline
0.835  & 1.571 $\pm 0.007 \pm 0.101 $ \\
\hline
0.917  &1.419 $\pm 0.009  \pm 0.132 $  \\
 \hline
0.986  & 1.341 $\pm 0.010 \pm 0.147 $ \\
\hline
1.088  & 1.272 $\pm 0.010 \pm 0.156 $ \\
\hline
1.167  & 1.121 $\pm 0.013 \pm 0.153 $ \\
\hline
1.261  & 0.955 $\pm 0.016 \pm 0.146 $ \\
\hline
1.384  & 0.874 $\pm 0.016 \pm 0.269 $ \\
\hline
 1.522 & 0.730 $\pm  0.012\pm 0.280  $ \\
\hline
1.645  & 0.708 $\pm 0.009 \pm 0.257 $ \\
\hline
1.795  & 0.617 $\pm 0.007 \pm 0.254 $ \\
\hline
1.967  & 0.581 $\pm 0.006 \pm 0.223 $ \\
\hline
 2.177 & 0.636 $\pm 0.003 \pm 0.187 $ \\
\hline
\end{tabular}
}

\caption{ Data on $\alpha_{g_1}(Q)$ from JLab experiments EG4 (left), EG4/E97110 (top right) and EG1dvcs (bottom  right).}
\label{tab:alpha}
\end{table}
%%%%%%%%%%%%%%%%%%%%%%%%%%
\FloatBarrier

\end{document}